\documentclass[twoside,fleqn]{article}
\usepackage{epsfig,espcrc2}

\newcommand{\eq}{\begin{equation}}

\newcommand{\eqx}{\end{equation}}

\newcommand{\eqn}{\begin{eqnarray}}

\newcommand{\eqnx}{\end{eqnarray}}

\newcommand{\cor}[1]{\left\langle{#1}\right\rangle}

\newcommand{\AmS}{{\protect\the\textfont2

  A\kern-.1667em\lower.5ex\hbox{M}\kern-.125emS}}

\title{Towards the lattice study of M-theory}

\author{R.A. Janik  and J. Wosiek \address{
M.Smoluchowski Institute of Physics, Jagellonian University, Cracow}
   \thanks{e-mail: wosiek@thrisc.if.uj.edu.pl,ufrjanik@jetta.if.uj.edu.pl}  }

\begin{document}

\begin{abstract}
We propose the Wilson discretization of the  supersymmetric
Yang-Mills Quantum Mechanics as a lattice version
of the matrix model of M-theory. An SU(2) model is studied numerically
in the quenched approximation for D=4. 
A clear signal for the existence of two different phases  
is found and the continuum pseudocritical
temperature is determined. We have also extracted the continuum limit of
the total size of the system in both phases and for different temperatures. 
\end{abstract}


\maketitle

\noindent 1. {\em Introduction.}
Since the discovery of the web of dualities linking different
superstring theories, it has been suggested that all these theories are
just different perturbative expansions of a single 11-dimensional
theory christened M-theory, The true nature of this theory remains,
however, mysterious. In \cite{BFSS} Banks, Fischler, Susskind and Shenker
proposed a concrete formulation for the degrees of freedom of M-theory
in a flat 11-dimensional background.

In this contribution we would like to report on the work done in \cite{US} on
a lattice formulation of the BFSS matrix model, thus allowing for a
nonperturbative treatment. We use the version with euclidean signature
suitable for studying thermodynamical properties of the theory. In
fact it turns out that there is a very rich spectrum of physics
involved. A study of these properties forms the main motivation for
our investigation. Below we briefly review the physics involved as 
it also provides an incentive for further study.

The BFSS lagrangian for finite $N$ is a dimensional reduction of 10D
SU(N) SYM down to 1 dimension, schematically
\begin{equation}
S\!\!=\!\!\int\! dt \left({1\over 2} 
(D_tX^i)^2+[X^i,X^j]^2+ fermions
\right)\!.
\label{QM}
\end{equation}

Equivalently it describes a system of $N$ D0 branes in the decoupling
limit. All the dynamics at finite temperature $T$ can be expressed in
terms of a dimensionless
coupling constant $g^2_{eff}=g^2_{YM}N/T^3$. At high temperatures the
system is perturbative. As we lower the temperature we enter the
strong coupling regime.

At this stage we may apply the framework of
the AdS/CFT correspondence (see \cite{IMSY,MAR} in this context). An
equivalent description is provided by a 10 dimensional Black Hole (BH)
supergravity solution, whose Hawking temperature is identified with
the temperature of the SYM quantum mechanics (\ref{QM}). In particular
the Bekenstein entropy of the black hole calculated from the area of
the horizon should be identified with the entropy of the SYM quantum
mechanics (for recent new results see \cite{KAB}) which thus gives a
microscopic realization of the `macroscopic' BH entropy. It would be very
interesting to directly observe this prediction by studying
numerically (\ref{QM}).

Another interesting point is the nature of the transition from the
perturbative to the strong coupling phase. In \cite{HOR} it has been
identified with the Horowitz-Polchinsky correspondence point between a
black hole phase and `excited string' phase. It's detailed nature is
still unknown.

As one lowers the temperature, one reaches a regime where the string
coupling becomes strong and one has to lift the 10D BH solution to a
11D `black wave'. Further down on the temperature scale the 11D `black
wave' localizes to a 11D BH.

All the above physics should be contained in the strong coupled regime
of (\ref{QM}). For most of these questions supersymmetry is
crucial. In our exploratory study we perform quenched simulations and
focus on showing the existence of two phases in the high and low temperature
regimes. When performing simulations in the full unquenched model the
observed phases should correspond to the black hole phase and the
perturbative phase. In the following we present the lattice
formulation, give details on the algorithms and present results of 
first exploratory quenched simulations for $N=2$.

\noindent 2. {\em Lattice formulation.}
A direct Monte-Carlo simulation of the action (\ref{QM}) is rather nontrivial. 
The potential for the bosonic fields $[X^i,X^j]^2$
has flat valleys which may be difficult to handle numerically. 
Furthermore it is essential to impose the Gauss law constraint or
equivalently to project onto gauge invariant states. This may be
important especially as the supersymmetry algebra in the continuum
closes only on the space of gauge invariant states.
Nevertheless the nonperturbative studies of the IIB matrix models
(SYM reduced to zero dimensions)
in this formulation are well advanced \cite{JAP,KOP}.

In order to overcome the above difficulties, and to use standard, 
well developed lattice techniques we decided, however, 
to follow a different route. 
The action (\ref{QM}) comes from the standard SYM action in $D=10$,
upon assuming that all fields do not depend on spatial coordinates. We
may now use the Wilsonian discretization in $D=10$ and impose the
constancy of link variables in the spatial directions. In this manner
we  obtain at once a $D$ dimensional hypercubic lattice 
$N_1\times\dots\times N_D$ reduced in all space directions to $N_i=1$,
$i=1\dots D-1$. 
Gauge and fermionic variables are assigned to links and sites of the new 
elongated lattice in the standard manner. The gauge part of the action 
is
\eq
S_G=-\beta
\sum_{m=1}^{N_t} \sum_{\mu>\nu}
{1\over N} Re( \mbox{\rm Tr} \, U_{\mu\nu}(m) ),
\label{SG}
\eqx 
with 
$\beta=2N/a^3 g^2,$  
and $U_{\mu\nu}(m)= U_{\nu}^{\dagger}(m)U_{\mu}^{\dagger}(m+\nu)
U_{\nu}(m+\mu)U_{\mu}(m) $, 
$U_{\mu}(m)=\exp{(iagA_{\mu}(a m))}$, where $a$ denotes the
lattice constant and $g$ is the gauge coupling in one dimension.
The integer time coordinate along the lattice is $m$.
Periodic boundary conditions $U_{\mu}(m+\nu)=U_{\mu}(m)$, $\nu=1\ldots
D-1$,
guarantee correct classical continuum limit (\ref{QM}) with $X_i=g A_i$.
In this formulation the projection on gauge invariant states 
is naturally implemented. Addjoint fermions can be included as in the 
lattice studies of SUSYM theory \cite{MM,US}.

\noindent 2. {\em Results.} 
We choose the distribution of the Polyakov line
$ P={1\over N}  \mbox{\rm Tr} \prod_{m=1}^{N_t} U_D(m),$
as an order parameter.
Similarly to lattice QCD, symmetric concentration of the eigenvalues
around 0 indicates a low temperature phase (which would have here the
interpretation of a black hole phase) where 
$\cor{P}\sim 0$,  while clustering around $\pm 1$ (for SU(2)) is
characteristic of the high temperature (elementary excitations) phase. 
\begin{figure}[htb]
\epsfig{width=7.5cm,file=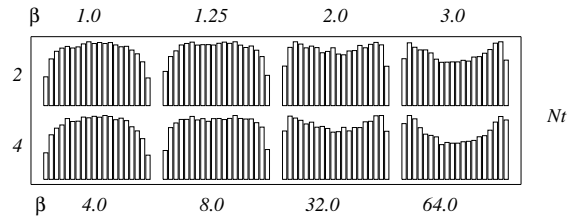}
\caption{Distribution of the Polyakov line $P$, $-1<P<1$, 
for different $\beta$ and $N_t$.
  }
\label{fig:f1}
\end{figure}    
Our first result is evident from Fig. 1.
For each $N_t$ a definite change of the shape with $\beta$ is
observed. 
The system (\ref{SG}) definitely shows 
the onset of the phase change, even in the quenched approximation
and for $N=2$. 
\begin{table}    
\begin{center}
\begin{tabular}{ccc} \hline\hline
$N_t$ &  $\beta_{low}$ & $\beta_{up}$  \\
\hline
2 &  1.25 & 1.5 \\
3 &  3.5  & 5.0 \\
4 &  8.0  & 16.0  \\ 
5 & 15.0  & 40.0  \\ 
\hline
\multicolumn{3}{c}{$fit:\;\;\;\;$     $\beta_c=\alpha N_t^{\gamma}$ }\\
\hline
$\chi^2/NDF$    &  $ \alpha $  & $\gamma$ \\
$.55/2 $       & $.17\pm .05$ & $3.02\pm .33$ \\
\hline\hline  
\end{tabular}
\end{center}
\caption{Estimated location of the transition region
$\beta_c\in(\beta_{low},\beta_{up})$ 
for different lattice sizes $N_t$ and results of the power fit.}
\end{table}
Second, the dependence of the  
pseudocritical temperature $\beta_c$ on the time extent $N_t$ is
consistent with the  
continuum limit expectations $T_c \sim (g^2 N)^{1/3}$ \cite{KAB}.
Indeed, the temperature of a system is given by $T=1/(a N_t)$ 
which implies $\beta_c \sim N_t^3$. This is confirmed in Table 1
where $\beta$ intervals where the 
transition occurs are presented for several lattice 
sizes $N_t$. Results of the 
power law fit are also quoted. A good quality of the fit and the agreement with
the canonical exponent, $\gamma=3$, is encouraging. Simultaneously, we
obtain the proportionality coefficient $\alpha$ which gives
\eq
  T_c=({\alpha\over 2N^2})^{1/3}(g^2 N)^{1/3}=(.28\pm.03) (g^2 N)^{1/3}. \label{TC}
\eqx
The coefficient in this relation 
has been determined for the first time. Only proportionality of the two
scales was considered \cite{HOR,MAR,KAB}. The 
pseudocritical temperature and $\alpha$ may depend on $N$, therefore  
similar analysis for higher gauge groups is necessary.

Next we study the total size of the system 
$R^2=g^2\sum_a (A_i^a)^2$ \cite{KAB}. We define for $SU(2)$ 
\eq
  \cor{R^2}\equiv \left(4-\cor{(\mbox{\rm Tr}(U_s))^2}\right)/a^2,
\label{rms} 
\eqx
where $U_s$ is any space link. 

    We used mostly the standard
local Metropolis update with enough thermalization and decorrelation sweeps
to take care of the critical slowing down. 
For example when running at $\tilde{a}=1.0$  we used
5000 thermalization and 50 decorrelation sweeps, while for 
$\tilde{a} \equiv a g^{2/3} =0.1$
about $10^6$ thermalization and 5000 decorrelation sweeps were required
to achieve independence of the  starting configuration.
This also agrees with the dynamical exponent $z=2$. In addition we have developed
the new heat bath algorithm
designed for an update 
of the space-space plaquettes in (\ref{SG}) which contain twice the same SU(2)
link. In the standard $S_3$ parametrization of SU(2) distribution of
the three vector $\vec{u}, |\vec{u}|=u \leq 1$ is the product of the three dimensional
gaussian and the cosh factor which depends on $\sqrt(1-u^2)$. At fixed  
values of the neighbouring links the center of the gaussian may lie outside
the kinematical region $u<1$, which even for intermediate $\beta$ makes 
the simple gaussian generation inefficient.
However by a suitable choice of the coordinate system\footnote{The one with the
third axis
pointing towards the maximum of a gaussian on a boundary sphere $u=1$.}
the problem becomes manageable. The quadratic part of the $\log{cosh}$ factor
is included in the above gaussian and deviations from the exact function
are included by the accept-reject step.     
Results obtained
with the new heat bath and the standard Metropolis agree within
statistical errors for $\beta < 64 $. For higher $\beta$ the accept-reject
step becomes inefficient.
 For higher $\beta$ we have also monitored the
correlation length in the torelon channel at zero 
temperature. It shows the canonical scaling with $a$ as expected.
\begin{figure}[htb]
\epsfig{width=7.5cm,file=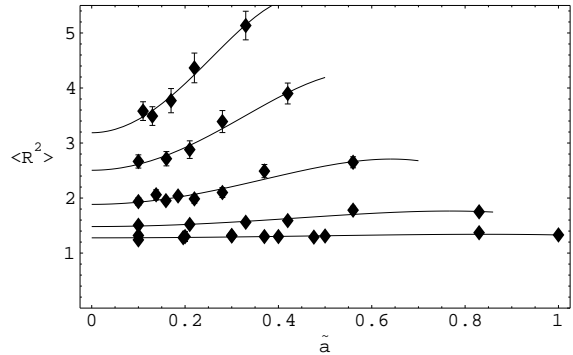}
\caption{ Dependence of the total size of the system on $a$, for
$\tilde{T}\equiv T g^{-2/3}=.1-.3,.6,.9,1.2$ and $1.5$ (upwards) 
in units of $g^{2/3}$. Quartic fits are represented by the solid lines. }
\label{fig:f2}
\end{figure}
Fig. 2 shows the 
dependence of $<R^2>$ (in units of $g^{2/3}$)  on $\tilde{a}$, for several values of the 
temperature $\tilde{T}$. 
MC results depend smoothly on $a$, at fixed T,  confirming
the existence of the continuum limit (\ref{rms}). The $a$ dependence
is   
different in low and high temperature regions. 
For $\tilde{T}>1.5 $ simulations for smaller $\tilde{a}$ are
required in order to see the quadratic approach to the continuum.
Practically the same results were obtained when we
extracted $\cor{R^2}$ from another lattice observable
$|Tr(U_s)|$.  
\begin{figure}[htb]
\epsfig{width=7.5cm,file=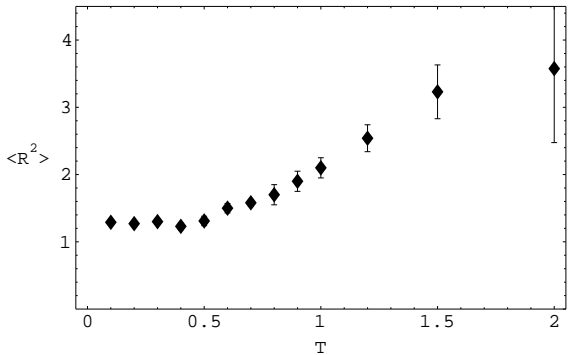}
\caption{ Size of the system (\ref{rms}) extrapolated to the continuum,
as a function of the temperature.}
\label{fig:f3}
\end{figure}
Fig. 3 shows the size of a system extrapolated
to $a=0$ as a function of the temperature. 
Both quadratic and quartic fits of $a$ dependence were used.  
The stability
of quadratic fits with respect to removing one or two data points with
smallest $a$ was also checked. Results of the extrapolation were
stable with respect to all these variations.
Displayed errors include above systematic effects. 
The location of the transition region
agrees approximately
with the estimate (\ref{TC}) of the pseudocritical temperature
$\tilde{T_c}=0.35\pm0.04$ 
Again, it is evident that the 
system is indeed different in the two regimes. Moreover, our results
agree qualitatively 
with the analytical prediction obtained from a gap equation 
in the infinite N limit \cite{KAB}. The latter gives a temperature 
independent constant at low temperatures and the classical $T^{1/2}$ 
growth for high temperatures. We have also found a reasonable agreement
with a simple mean field model for $SU(2)$ with the gauge
projection which will be discussed elsewhere.

\noindent {\em 3. Outlook.}
We have constructed, for the first time, a matrix model of M-theory
on a lattice.
MC simulations show that even the strongly simplified version
of the model has different high and low temperature phases as 
expected in the original
theory. Pseudocritical temperature and total size of the system
show canonical approach to the continuum limit. This indicates that,
at least in this respect, the system may be simpler than the ones 
encountered in lattice studies of field theories in extended space-time.
In particular, restoration of the supersymmetry,
broken by lattice discretization, is feasible. 

Of course results
presented here are only a hint that we may be on a right track.
One urgent task is to repeat present programme for higher N and D=10.
This, although relatively simple, may require some refinement of existing
algorithms or inventing new ones dedicated to the linear systems.
Equally important is to include dynamical, supersymmeric fermions.
For D=4 this is again relatively standard and would allow us to study
full intricacies of lattice breaking and subsequent 
restoration of the supersymmetry in the continuum limit. 
For D=10 the problem of dynamical fermions is open and presents
an exciting challenge. Again the linear nature of the system may offer
some simplifications. 

    To conclude, we believe that many interesting applications became
available opening new possibilities for the nonperturbative studies of
the prototypes of M-theory.

\noindent {\em Acknowledgements.} J.W. thanks D. Maison and 
P. Breitenlohner for a discussion. This work is  supported by the
Polish Committee for Scientific Research under grants no. PB
2P03B00814 and PB 2P03B01917.

\end{document}